\documentclass[journal=aanmf6,manuscript=article]{achemso}
\usepackage[utf8]{inputenc}

\usepackage{achemso}
\usepackage{url}

\usepackage{amsmath}
\usepackage{tabularx}
\usepackage{graphicx}
\usepackage{subfig}
\usepackage{multirow}
\usepackage{makecell}
\usepackage{upquote}
\usepackage[normalem]{ulem}
\usepackage{physics}

\newcolumntype{Y}{>{\raggedleft\arraybackslash}X}
\newcolumntype{Z}{>{\centering\arraybackslash}X}

\newcommand{\refSIPhases}{S1}
\newcommand{\refSIIntercBonds}{S2}

\newcommand{\refSIxzRelaxYShear}{S6}

\newcommand{\refSILoad}{S8}
\newcommand{\refSISymmetries}{S9}

\title{Atomistic Mechanisms of Sliding in Few-Layer and Bulk Doped MoS$_2$}

\author{Enrique Guerrero$^\dagger$}
\email{eguerrero23@ucmerced.edu}
\affiliation{$^\dagger$Department of Physics, University of California, Merced, Merced, CA 95343}

\author{David A. Strubbe$^\dagger$} \email{dstrubbe@ucmerced.edu}
\affiliation{$^\dagger$Department of Physics, University of California, Merced, Merced, CA 95343}

\begin{document}

\maketitle

\begin{abstract}
Sliding of two-dimensional materials is critical for their application as solid lubricants for space, and also relevant for strain engineering and device fabrication. Dopants such as Ni surprisingly improve lubrication in MoS$_2$, despite formation of interlayer bonds by intercalated Ni, and the mechanism has remained unclear. While sliding on the atomistic level has been theoretically investigated in pristine 2D materials, there has been little work on doped forms, especially for the complicated case of intercalation. We use density functional theory to study sliding of Ni-doped MoS$_2$, considering Mo/S substitution and octahedral/tetrahedral intercalation. We find that bulk and trilayers are well described by pairwise bilayer interactions. Tetrahedral intercalation between layers dramatically increases their sliding barrier, but minimally affects sliding between adjacent undoped layers, thus preserving effective lubrication. We provide an atomistic view of how sliding occurs in doped transition-metal dichalcogenides, and a general methodology to analyze doped sliding.
\end{abstract}

MoS$_2$ is a semiconducting transition metal dichalcogenide (TMD) with low-friction lubricative properties due to the ease of shearing of its lamellar structure,\cite{Vazirisereshk} and is often used for space applications.\cite{Krause,Babuska} Sliding can also be important for strain engineering,\cite{Pena} nanoelectromechanical systems,\cite{LeeNanoElec} oscillators\cite{YeOscil}, sliding ferroelectrics,\cite{WuFerro} and assembly of twisted bilayer structures.\cite{Liao} For all these applications, it is important to understand how the atomistic mechanisms of slip which are understood in bulk materials are modified on the nanoscale for sliding in novel 2D materials, and especially how dopants influence sliding. MoS$_2$ can be doped to tune the electronic properties,\cite{Bernardi} increase the catalytic activity,\cite{Mundotiya, Mao} improve lubrication,\cite{Vazirisereshk, Vellore}, or change the layer dependence of friction.\cite{Acikgoz} Friction in two-dimensional (2D) materials like MoS$_2$,\cite{Levita, Yang, Hu, LWang, Liang} graphene,\cite{Ta, LWang} and hexagonal boron-nitride (hBN)\cite{Marom, Cheng} has been simulated using models of varying complexity, but computational studies of doped materials are rare, so far only for intercalated graphite \cite{Pang} and substituted hBN \cite{Cheng}.
Sliding in doped TMDs, doped bulk 2D materials, or transition-metal-doped 2D materials does not seem to have been studied computationally. The usual approach for exploring sliding in 2D materials is through the sliding potential-energy surface:\cite{Ustunel} energies are computed as a function of sliding displacement, and and the effect of variables such as load or layer orientation can be examined. In these van der Waals (vdW) mediated systems, a consensus has formed that the interfacial geometries and resulting charge density fluctuations are primarily responsible for the sliding potential's shape\cite{Levita, JWang, LWang, Cheng}. This idea has been used as the basis for a classical approximation of the potential.\cite{Blumberg, Marom}
The sliding potential has been shown to be altered by S corrugation from layer rotation\cite{Levita} or introduction of vacancies,\cite{JWang} which can also happen by formation of distorted 1T phases.\cite{KarkeePanoply} 
This picture based on weak vdW interactions could be altered when there are dopants between layers,\cite{ChenPhaseTrans,Yang} such as in intercalation of Ni\cite{Guerrero} or Re\cite{Guerrero_Re} where the dopants form covalent interlayer bonds. Dopants, especially intercalated dopants, introduce local bond strain and atomic mobility which must be carefully considered during computation and analysis when compared to the standard approach. Ni increases the degrees of freedom, meaning we have to balance relaxing the local structure and preserving a quantifiable sliding amount. Previous simplifying methods like the registry index\cite{Blumberg} are not easily applicable with intercalation where significant structural changes (e.g. bond changes) can occur.
Finally, choices in the initialization of structures in independent computations can affect the final result and lead to pathway discontinuities.

Previous sliding studies have extended to bilayers with vacancies\cite{JWang, ZGYu}. Like dopants, vacancies could introduce local bond reorganization which must be considered. Vacancy complexes as large as seven atoms were found to universally reduce the sliding barrier while leaving the potential's shape unaffected.\cite{JWang} Dissociated H$_2$O's between layers has been found to severely diminish lubrication.\cite{Yang}
%
%
Conversely, Cheng et al. consider carbon doping in hexagonal boron-nitride bilayers with DFT. C's presence was shown to decrease the friction for substitution of either B or N\cite{Cheng}---the interlayer interaction is weakened by the dopant and thus lubricity is improved. By contrast, calculations showed that Li intercalation in graphite greatly increased barriers to sliding, due to electrostatic interactions after charge transfer from Li to semimetallic graphene layers.\cite{Pang} We have previously computed that Ni-doped MoS$_2$ does not reduce the binding energy between adjacent layers as Li might do\cite{EnyashinLi} but rather increases it substantially in t-intercalated doping,\cite{Karkee} opposite to what was found for C-doped BN. While intercalated Li can be thought of like ball bearings that decouple MoS$_2$ layers and would reduce friction (though surprisingly it is not clear this effect has been measured), intercalated Ni instead increases the coupling between MoS$_2$ by bonding. Ni has been found to not significantly alter MoS$_2$'s friction and enhance its resistance to wear, or layer flaking,\cite{Stupp, Vellore} and may have promise for increasing the capabilities of lubrication in space.\cite{Krause} This is consistent with the idea that Ni-doping improves MoS$_2$'s resistance to wear (material loss through flaking) but appears counter to the idea that Ni doping improves lubrication.\cite{Stupp, Vellore} Calculations of Graphene/Fe$_2$O$_3$ interfaces with B, P, Si, and S dopants have found a frictional reduction. These defective cases do not consider the possibility of geometrical reorganization during sliding beyond the out-of-plane direction.\cite{Levita} While that direction is likely the most energetically important, local bond distortions could occur at the defect sites during sliding and alter the sliding potential.

\begin{figure}
	\includegraphics[width=325px]{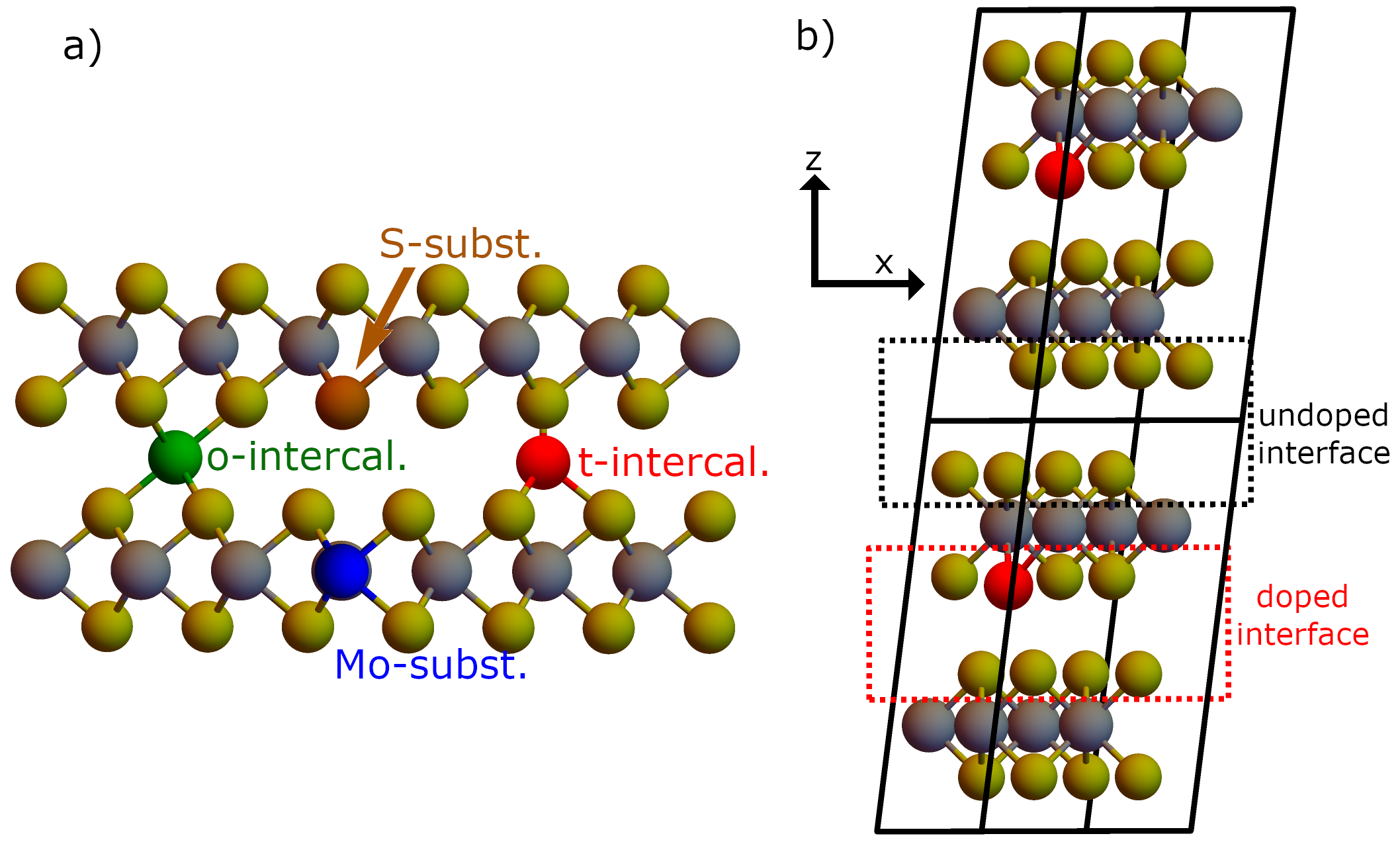}
	\caption{a) The four dopant sites considered in this work, previously found to be stable or metastable.\cite{Guerrero} Bilayer and bulk structures with similar atomic coordinates are computed. b) A sheared S-substituted cell showing two distinct interfaces in each bulk cell.}
	\label{fig:DopantSites}
\end{figure}

%
%
%
%
Ni has been found to improve MoS$_2$'s lubrication performance\cite{Stupp}, enhance MoS$_2$'s catalytic activity, and alter the electronic properties.\cite{Guerrero} 
We have previously computed four potential Ni dopant sites to MoS$_2$\cite{Guerrero, Karkee}: Mo substitution, S substitution, tetrahedral (t-) intercalation and octahedral (o-) intercalation and consider all of them in this work. 
In this article, we use density functional theory to explore how the presence of Ni modifies potential-energy surfaces and structures during sliding under different atomic constraints. We consider the Ni dopant locations in Fig. \ref{fig:DopantSites} and find the relation between bilayer sliding and bulk shearing. We provide insight and methods for analysis generally applicable to sliding of doped 2D materials.

We begin with doped and pristine 2H-MoS$_2$, in the AA$'$ stacking.\cite{Blumberg} Initial structures are generated by fully relaxing doped bulk structures, then forcing hexagonal symmetry onto the cell (relaxed cell angles differ from the pristine $120^\circ$ by at most $0.025^\circ$, and energies differ by $10^{-4}$ eV/atom). Our coordinate system has the $x$-axis in the zigzag direction and $y$-axis in the armchair direction. The top layer is then displaced in the $x$-direction by $\Delta x$ in steps of 1/12th of the cells' respective lattice constant $a$. $a$ was set to 3.19018 \AA, 6.39945 \AA, 6.38666 \AA, 6.35100 \AA, and 6.39413 \AA\ for pristine, Mo-substituted, o-intercalated, S-substituted, and t-intercalated in both bulk and bilayers. In bulk, the $x$-component of the out-of-plane vector $c_x$ was set to $2\Delta x$, such that every pair of layers experienced relative sliding of $\Delta x$. In intercalation, Ni was moved along with the mobile top layer, since it was closer to the intercalant. Substitutional Ni was moved along with the layer it is in. Energies are computed by a self-consistent field (SCF) computation and then the structure is relaxed by three successive levels of constraints as summarized in Fig. \refSIPhases. (1) Atoms are allowed to relax in the $x$-direction and the $c_{z}$ lattice parameter is relaxed ($z$-relax). (2) Atoms are further relaxed in their $y$-coordinates ($yz$-relax). (3) Atoms are further still relaxed in $x$ ($xyz$-relax). Throughout stages (1)-(3), the Ni atom is allowed to relax in all coordinates so as to capture its preferred movement throughout sliding.

Each constraint gives us different information about the structure throughout sliding. $z$-relax yields the 1D-sliding potential as it relates to interlayer distances. $yz$-relax allows us to access the 2D-sliding potential and find deviations from the low-barrier sliding pathway. $xyz$-relax allows us to test whether shearing occurs like a deck of cards or with the presence of slip planes. Furthermore, the initial structure after sliding but before relaxation can be thought of as stage (0), no relaxation.
A similar data set was constructed for $y$-direction sliding and can be found in the Supplementary Information.

It is important to relax the Ni atom's position during intercalated sliding because the local geometry is expected to change when it experiences different local environments. Ni bonding remains at 6 and 3 bonds for Mo-substituted and S-substituted sites throughout sliding respectively, but bonds form and break for intercalated sliding. Given full relaxation freedom in $xyz$-relax, both structures follow paths with four Ni-bonds, indicating the high stability of that bond configuration.

The sliding potential in both bulk and bilayer shows a mirror symmetry in the $x$-direction. This is because the AA$'$ stacking of pristine, Mo-substituted, S-substituted, o-intercalated, and t-intercalated belongs to symmetry point groups D$_{\rm 6h}$, D$_{\rm 3h}$, C$_{\rm 3v}$, C$_{\rm 3v}$, and C$_{\rm 3v}$ respectively, which include this mirror symmetry. Moreover, interlayer sliding in the $x$ direction transforms as the function $xz$, which in each point group belongs to an irreducible representation other than the totally symmetric one; therefore the forces are zero along the sliding direction at AA$'$ and other high-symmetry points. 
In pristine, this information can be used to reduce the computational workload, but once the doped structure is relaxed, there is slight symmetry-breaking in all doped cases, quantified in Fig. \refSISymmetries. Since every interface involves at least one pristine layer, the symmetry of the full sliding potential shows hexagonal symmetry.

\begin{figure}
	\includegraphics[width=450px]{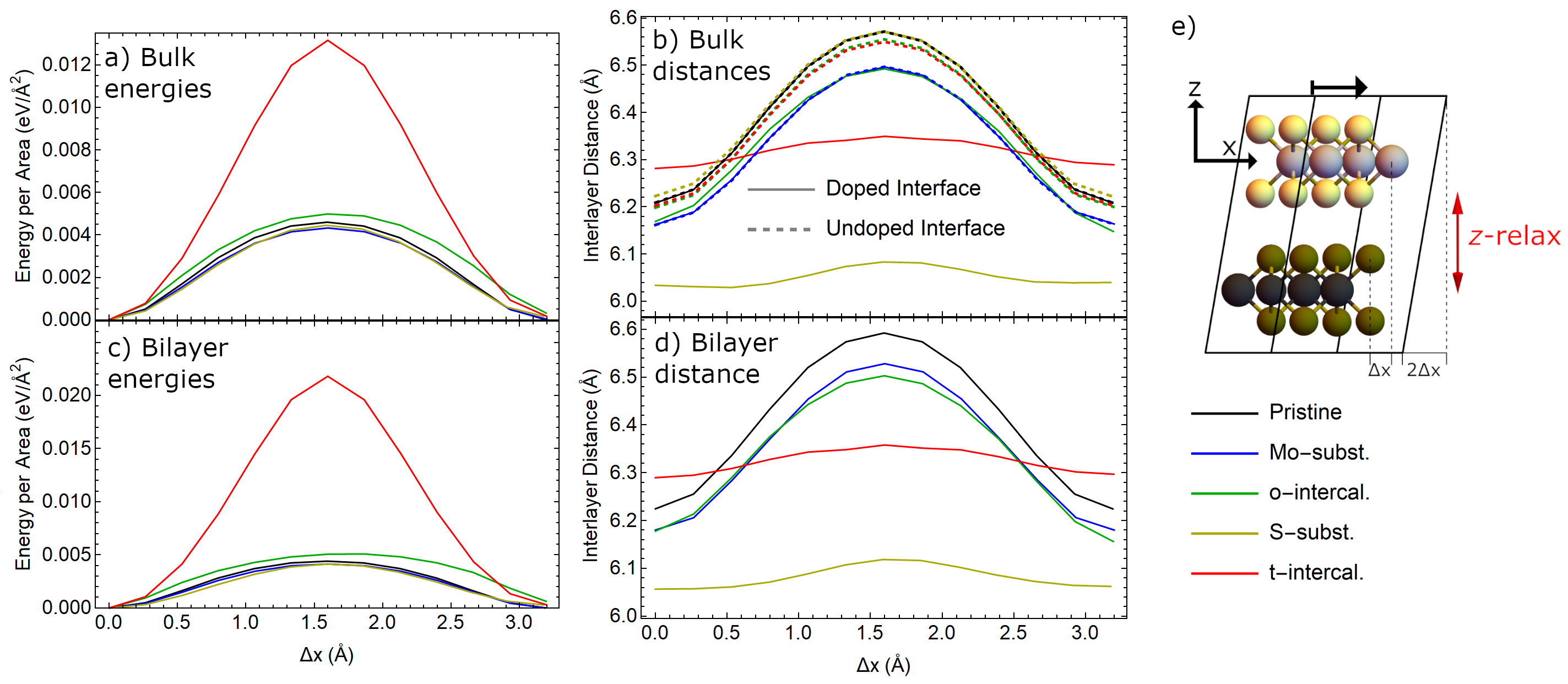}
	\caption{Energies and interlayer distances for $x$-sliding after $z$-relax conditions were applied as in e). Energies for a) bulk and c) bilayer shown are divided by the areas of the actively sliding interfaces (i.e. 2$\times$ cell's area in bulk, once in bilayer). Interlayer distances in b) are compared between doped (solid) and undoped (dashed) interfaces. S-substitution and t-intercalation show the largest difference between the two, where they have a strong preference to keep the interlayer distance the same. d) Bilayer distances are similar in shape to bulk. }
	\label{fig:zRelax}
\end{figure}

Results of $z$-relax in Fig. \ref{fig:zRelax} show that the sliding barrier is massively increased in t-intercalation, unlike previous results with vacancies.\cite{JWang} As layers slide, the identifiers ``tetrahedral'' and ``octahedral'' are used to describe the initial structure prior to sliding, despite the fact that bond count and orientation change. 
The difference between t- and o-intercalations' energetic response to 1D-sliding can best be explained by changes in their bonding, (See Fig. \refSIIntercBonds) where o-intercalation is allowed to move into the more stable 4-bonded configuration while t-intercalation is forced to increase its bonds to 5 or 6. 
Fig. \ref{fig:zRelax}b separates $c_z$ into the two interlayer distances present in a cell. These distances are the same for pristine and Mo-substitution, where every other interface is equivalent, but are not for the other structures. This leads to the surprising result that t-intercalated and S-substituted structures attempt to keep their interlayer distance where the Ni is between two layers. This propensity for the interface in these sites to remain similar throughout sliding suggests that the slip planes occur away from the dopant site.

\begin{figure}
	\includegraphics[width=450px]{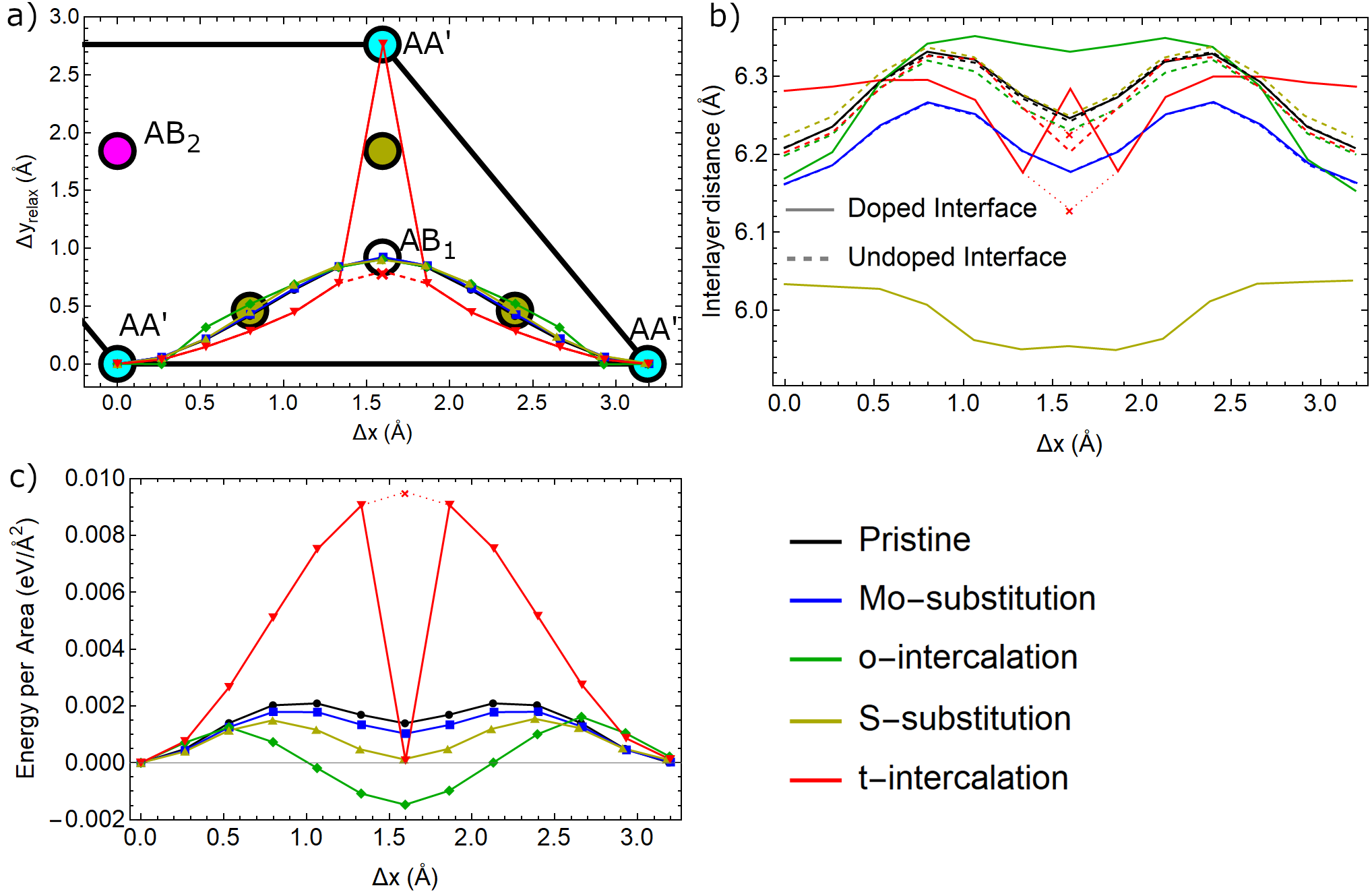}
	\caption{Structure and energy of bulk sliding layers by $\Delta x$ then relaxing in $y$ and $z$ coordinates. a) The relative displacement after relaxation. Colored circles indicate high-symmetry stackings as labeled, except that the yellow/brown circles mark the locations of energy extrema for S-substitution.
	Given the opportunity to relax in $y$, structures traversed the potential surface between stable and metastable stackings. b) Mo-Mo interlayer distances are shown for both doped and undoped interfaces. S-substitution, unlike others, does not increase its doped interface's interlayer distance, indicating an attraction between the dopant and its opposite layer. c) Energies along the sliding path are displayed as well. For b)-c), a red $\times$ indicates the t-intercalated structure as generated by a different scheme than other structures to target the AB$_1$ stacking.}
	\label{fig:yzRelax}
\end{figure}

In $y$-relax, the structures are free to follow a curved path indicative of the familiar zigzag sliding pattern\cite{Levita, Blumberg} of hexagonal lattices. This curved path is quantified as $\Delta y$, the mean difference in Mo $y$-coordinates when compared to the $\Delta x=0$ \AA\ structure after relaxation. The white circle in Fig. \ref{fig:yzRelax} represents the metastable AB$_1$ (Mo eclipses Mo) stacking of the pristine structures. Besides t-intercalation, doped structures follow a pathway to equivalent sites through the same metastable site. The t-intercalated structure instead relaxed to a stacking equivalent to the lowest-energy stacking, causing discontinuities in $\Delta y$ at $\Delta x=1.6$ \AA\ in Figs. \ref{fig:yzRelax} and \refSIIntercBonds. This is likely due to the deep well at the equivalent stable stacking to attract our system to that configuration, and small barriers could be overcome by the BFGS quasi-Newton algorithm as implemented in ESPRESSO.\cite{Fletcher} In Fig. \ref{fig:zRelax}, we included a structure computation which follows the trends better, but was obtained by altering the relaxation scheme. It was initialized by sliding a nearby structure, rather than sliding the initial AA$'$ structure. Furthermore, both intercalated structures deviate slightly in their sliding pathway due to forces from the Ni-S bonds. The results of this method are reminiscent of what may be achieved through the nudged elastic band method,\cite{Yang, Zarkevich} though we do not require the images to remain close by. This method accesses the 2D-sliding path by simply allowing the system to `fall' into low-energy pathways in the potential energy surface. Given the complexity introduced by local atomic organization, this analysis approach can capture the concerted relative sliding of layers but also allowing for local symmetry breaking with relaxations of atoms nearby the Ni atom. Along the zigzag sliding path, S-substituted layers attract the opposite layer during sliding (Fig. \ref{fig:yzRelax}b), as we also found when applying increasing loads in Fig. \refSILoad. For $y$-sliding in Fig. \refSIxzRelaxYShear, we instead use $xz$-relax conditions, but we found that the structure did not deviate from a straight path along the $y$-axis due to mirror symmetry in the $xz$-plane which is preserved for displacement along the $y$-axis.

\begin{figure}
	\includegraphics[width=450px]{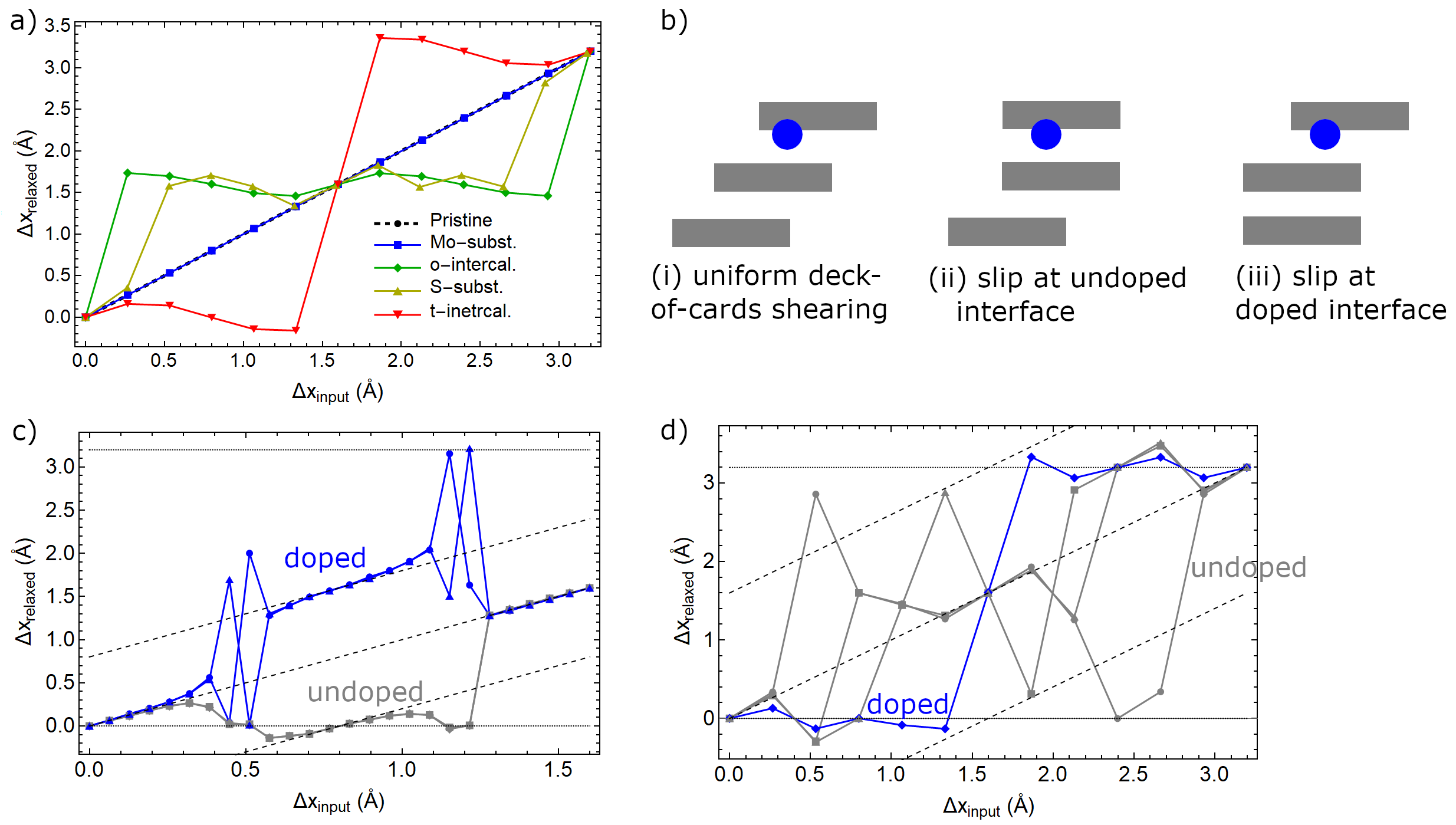}
	\caption{Geometry after bulk shearing in $x$ by $\Delta x$ then relaxing in $x$, $y$ and $z$ coordinates. a) The relaxed sliding distance $\Delta x_{\rm relax}$ as a function the input sliding distance $\Delta x$ with 2 layers per cell. b) A schematic of potential multilayer slip-plane configurations. c) Mo-substituted and d) t-intercalated interlayer $x$ displacements, $\Delta x ({\rm relax})$ for doped (blue) or undoped (gray) interfaces in bulk structures with four layers per cell.}
	\label{fig:xyzRelax}
\end{figure}

We use the least restrictive constraint, $xyz$-relax, to detect preferred locations of slip-planes. Fig. \ref{fig:xyzRelax}a shows $\Delta x_{\rm relax}$, the average difference between $x$-coordinates of Mo atoms of the two relaxed layers in the cell with respect to $\Delta x$ with AA$'$ stacking. By sliding the layers uniformly and shearing the cell then relaxing along that same sliding direction, we can see whether it is favored to have uniform relative displacements between layers, more displacement between doped interfaces, or less displacement between doped interfaces. For pristine and Mo substitution, the two interfaces are identical by symmetry and must have the same displacement, but other structures break this symmetry and can have different displacements. The t-intercalated interface shows a strong preference for the AA$'$ stacking. Conversely, S-substituted and o-intercalated interfaces prefer the metastable stacking. This suggests that these structures show a preference to the slip plane shown in \ref{fig:xyzRelax}b, but larger supercells in the $z$ direction are required to break the Mo-substituted symmetry and test its slip-plane preference directly.

To that end, we compute a few bulk supercells with 4-layers but only one doped layer in (Fig. \ref{fig:xyzRelax}c-d) (note this has half the Ni concentration of our structures thus far). Both Mo-substitution and t-intercalation relax to alternative stacking configurations rather than the input uniform sliding. We compute the $\Delta x_{\rm relax}$ of the four interfaces, or pairs of layers, and contrast the behavior of interfaces involving \textit{vs.} not involving the dopant. We find doped Mo-interfaces absorb the sliding, while the undoped interfaces attempt to remain in AA$'$ stacking. At around 0.5 and 1.2 \AA, instabilities due to small energy differences in layer orientation emerge, but the undoped interface ultimately remains closer to AA$'$ stacking. Conversely in t-intercalation, $\Delta x_{\rm relax}$ stays near 0 \AA\ throughout sliding for the doped-interface, just as in the two-layer per cell structure in Fig. \ref{fig:xyzRelax}a. Pristine structures, even when perturbed, relax to a uniform shearing as in Fig. \ref{fig:xyzRelax}b. This study locates the slip plane at the doped interface for Mo-substitution (as in Fig. \ref{fig:xyzRelax}b iii) and away from the interface at t-intercalation (as in Fig. \ref{fig:xyzRelax}b ii). The in-plane lattice is not relaxed.

\begin{figure}
	\includegraphics[width=450px]{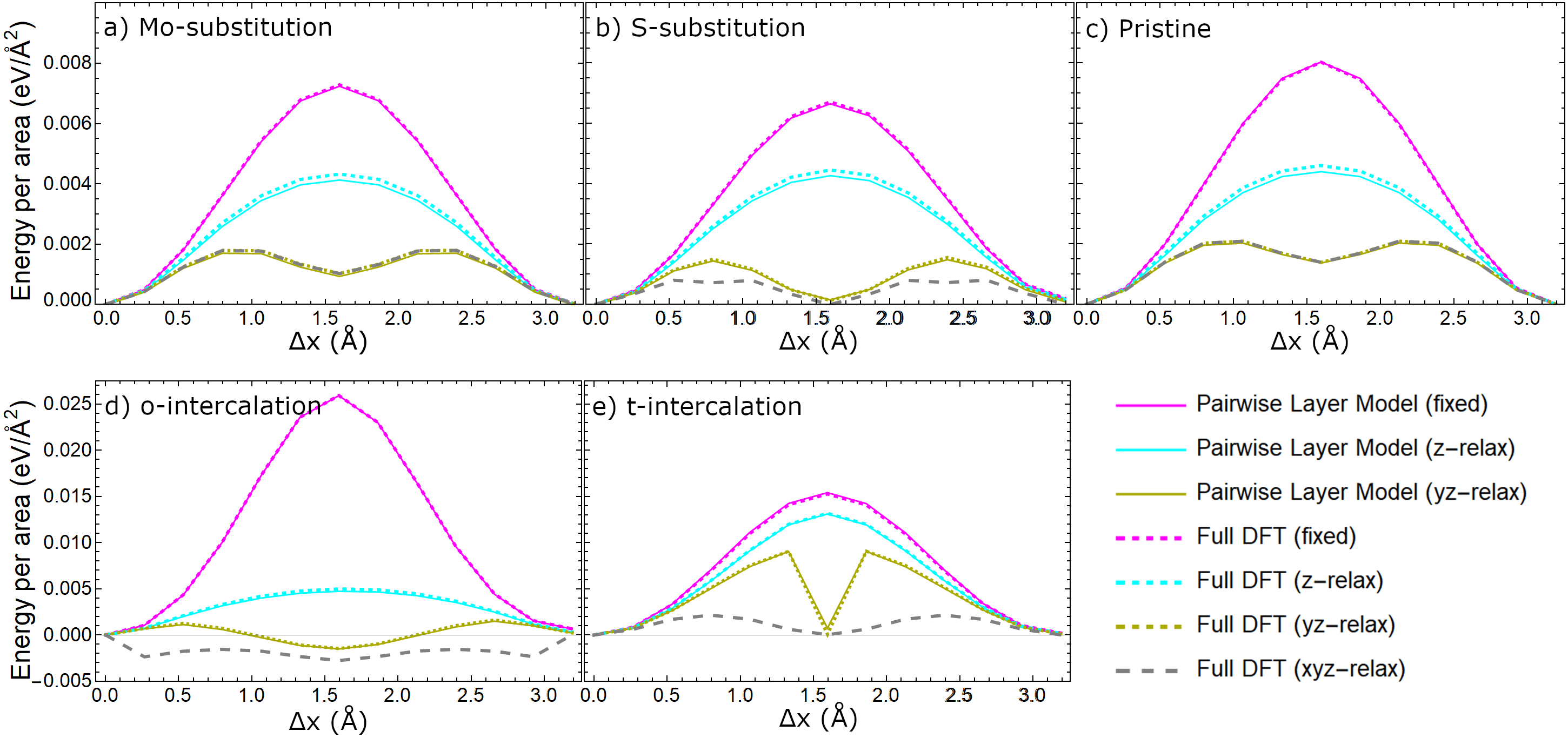}
	\caption{The computed (dashed) variation in sliding potential in bulk plotted against sums of bilayer sliding potentials (solid). Bilayer structures whose interfaces match those in bulk are used: two Mo-substituted/undoped bilayer interactions generate one bulk Mo-substituted sliding potential, others are sums of one undoped interface (i.e. pristine bilayer sliding) and one doped interface. The approximation would only be suitable for $xyz$-relax conditions if a full 2D sliding potential is computed, since relaxing bilayers in all coordinates yields AA$'$ stacking.}
	\label{fig:ConstructedShear}
\end{figure}

The bulk structure we have shown thus far contains two interfaces within each cell and have a concentration of one Ni per 8 MoS$_2$ units. To understand the range of interactions and quantify how well this may match other concentrations of dopants, we computed an isolated bilayer system with the same initial atomic coordinates as in bulk, but with vacuum in the $c$ direction. Relaxing such structures with identical constraints yields the same stacking configurations found in the bulk. This leads to the intuitive result that the interlayer interactions in the bulk structure can be computed as pairwise interactions of bilayers as summarized in Fig. \ref{fig:ConstructedShear}---effects are local to the layers, van der Waals and covalent, not modified by farther away or electrostatic charge transfer. This coincides with comments that the interfacial geometry,\cite{LWang} and not any deeper structures, are almost entirely responsible for the shape of the sliding potential. As shown in Fig. \ref{fig:ConstructedShear}, the practical result is that we can construct the bulk sliding energies by summing up appropriate sliding potentials of bilayer computations. Namely, bulk o-intercalation, t-intercalation, and S-substitution potentials are sums of their bilayer counterparts and the pristine computation while bulk Mo-substituted and pristine are twice their bilayer counterpart potentials. Fig. \ref{fig:ConstructedShear}d, for example, has two interface interactions in our bulk structures, one interface with no Ni atom and one with the Ni atom---so we approximate it by summing up a pristine bilayer sliding potential and the t-intercalated bilayer sliding potential to remarkable accuracy. In effect, this leads to the conclusion that arbitrary layered structures can be computed as an appropriate ensemble of bilayer computations that match the interfacial geometry. Work by Hu \textit{et al.}\cite{Hu} studied sliding in 5-layer systems using classical potentials. Such a study would be expensive using density functional theory due to the large number of atoms, but could be reduced in cost dramatically while achieving DFT-level accuracy by careful combination of four bilayer computations representing the interfaces.

To directly test the localization of the interaction at the interface, we computed a few trilayer scenarios. The top two layers were pristine in all scenarios. The third layer is initially attached to the middle layer and is set to either t-intercalated, Mo-substituted, or pristine. We treat the system up to $yz$-relax only. We find that the bottom pair of layers remain in the AA$'$ stacking throughout sliding in all conditions. The top pair of layers follow an identical trajectory as the pristine bilayer and the sliding potential per area is indistinguishable in all three scenarios, corroborating that only the active interface contributes to the sliding potential of 2D materials. As a lubricant, Ni-doped MoS$_2$ shows higher resistance to wear and lower friction than pristine MoS$_2$. Given the increased sliding potential maximum of t-intercalated MoS$_2$ when compared to pristine, one may be tempted to assume the increased lubrication incompatible with t-intercalation. Our result subverts that expectation. With previous computations of the dissociation energy,\cite{Karkee} t-intercalation may bind the layers therefore decreasing the wear rate. Slip in t-intercalated MoS$_2$ can occur at the undoped interface, avoiding high-barrier t-intercalated sliding and relegating all of the sliding to undoped interfaces.

This work elucidates the atomistic mechanisms of doped 2D materials and outlines a systematic pathway to study systems with otherwise large degrees of freedom. The analysis pathway used is applicable to any layered material and at computational levels other than DFT, though an adequate model of the vdW interaction is critical for accuracy. Given that the pairwise bilayer interactions accurately describe the bulk, arbitrarily-layered systems could be discretized and computed as bilayers that represent each interface. This could reduce computational cost while keeping accuracy of many-layered studies.\cite{Hu} The variable constraint method used here can be used to simplify the complicated structures with dynamic internal components, such as in S-substituted or intercalations. We used these to find a low-barrier sliding path, compare the Ni's site-dependent alterations to the sliding potential, and locate the likely slip planes in those structures. A potentially perplexing result that t-intercalation has a large sliding potential and yet can lead experimentally to lower friction can be explained as the presence of shear planes away from intercalated layers. Our results provide general insight into sliding of doped 2D materials, for use in applications and for tuning sliding interactions in 2D materials, and our work offers a general approach for analysis, including for mapping out sliding potential energy in 2D.

\section*{Methods}
We use the plane-wave DFT code Quantum ESPRESSO\cite{QE} version 6.6 to compute the properties of bilayer and bulk Ni-doped MoS$_2$, as in our previous work.\cite{Guerrero, Karkee} We use the Perdew-Burke-Ernzerhof (PBE)\cite{PBE} exchange-correlation functional and Grimme-D2\cite{GD2} (GD2) vdW corrections to capture the weak interactions which mediate the sliding potential in pristine MoS$_2$. We use optimized norm-conserving Vanderbilt (ONCV) pseudopotentials\cite{SG15} as parameterized by PseudoDojo.\cite{PseudoDojo}

$2\times2\times1$ supercells of the 6-atom unit cell were used with one Ni atom per cell. Half-shifted Monkhorst-Pack $k$-grids were used, with in-plane sampling of $6\times6$ for pristine $4\times4$ for doped cells, and an out-of-plane grid size of 2 for bulk and 1 for single or multi-layer structures. Kinetic energy cutoffs of 60 Ry were used. Self-consistent field (SCF) energies were computed to thresholds of $10^{-8}$ Ry. Total energies and forces were relaxed to $10^{-6}$ Ry and $10^{-4}$ Ry/Bohr. Stress relaxations used thresholds of 0.05 kbar. 

Bilayer structures were calculated with a fixed out-of-plane lattice parameter $c$ = 25 \AA\ to provide vacuum. Trilayers of pristine and doped structures were constructed by adding a third, pristine layer to the bilayer structure and adding the layer thickness to $c$. Four-layer out-of-plane supercell bulk structures are extensions of the pristine, Mo-substituted, and t-intercalated bulk structure cells with the dopant kept in only one layer.

The number of bonds to a Ni atom $i$ from neighbors $j$ is defined using the Tersoff bond order function\cite{Tersoff} $f_C(r)$, where $r$ is the interatomic distance for a given pair of atoms:
\begin{equation}
	N_{i}^{\rm bonds} = \sum_{j \ne i}^{\rm atoms} f_C \left( | \vec{r}_i - \vec{r}_j | \right) =
	\begin{cases}
		1 & r\leq R-D \\
		\frac{1}{2}-\frac{1}{2}\sin(\frac{\pi}{2}\frac{r-R}{D}) & R-D<r<R+D \\
		0 & r\leq R+D \\
	\end{cases}
\end{equation}
where the bond cutoff $R=3.2$ \AA\ and the smoothing parameter $D=0.3$ \AA\ were chosen such that: (1) equilibrium o- and t-intercalated structures have 6 and 4 bonds respectively, (2) equilibrium Mo-substituted and S-substituted structures have 6 Ni-S and 3 Ni-Mo bonds respectively, and (3) variation in numbers of bonds with sliding is reasonably smooth. At equilibrium, the Ni-S bond lengths are 2.38 \AA, $2.12-2.17$ \AA, and 2.36 \AA\ for Mo-substituted, t-intercalated and o-intercalated respectively, while the Ni-Mo bond lengths in S-substitution are $2.68-2.78$\ \AA.\cite{Guerrero} 

\begin{acknowledgement}
	This work was supported by the Merced nAnomaterials Center for Energy and Sensing (MACES), a NASA-funded research and education center, under award NNH18ZHA008CMIROG6R. This work used computational resources from the Multi-Environment Computer for Exploration and Discovery (MERCED) cluster at UC Merced, funded by National Science Foundation Grant No. ACI-1429783, and the National Energy Research Scientific Computing Center (NERSC), a U.S. Department of Energy Office of Science User Facility operated under Contract No. DE-AC02-05CH11231.
\end{acknowledgement}

\begin{suppinfo}
 Summaries of geometries and energies of $y$-sliding in bulk and bilayer, Ni bonding during sliding, analysis of one interfaces in trilayers, analysis of load at the AA$'$ stacking, and symmetry breaking (PDF).
\end{suppinfo}


\providecommand{\latin}[1]{#1}
\makeatletter
\providecommand{\doi}
  {\begingroup\let\do\@makeother\dospecials
  \catcode`\{=1 \catcode`\}=2 \doi@aux}
\providecommand{\doi@aux}[1]{\endgroup\texttt{#1}}
\makeatother
\providecommand*\mcitethebibliography{\thebibliography}
\csname @ifundefined\endcsname{endmcitethebibliography}
  {\let\endmcitethebibliography\endthebibliography}{}

\end{document}